**Title:** Identification of network motifs capable of frequency-tunable and robust oscillation

**Authors:** Matthew Bailey and Jaewook Joo

**Affiliation**: Department of Physics and Astronomy, University of Tennessee, TN 37996-1200

**Abstract:** Oscillation has an important role in bio-dynamical systems such as circadian rhythms and eukaryotic cell cycle. John Tyson *et. al.* in Nature Review Mol Cell Biol 2008 examined a limited number of network topologies consisting of three nodes and four or fewer edges and identified the network design principles of biochemical oscillations. Tsai *et. al.* in Science 2008 studied three different network motifs, namely a negative feedback loop, coupled negative feedback loops, and coupled positive and negative feedback loops, and found that the interconnected positive and negative feedback loops are capable of generating frequency-tunable oscillations. We enumerate 249 topologically unique network architectures consisting of three nodes and at least three cyclic inhibitory edges, and identify network architectural commonalities among three functional groups: (1) most frequency-tunable yet less robust oscillators, (2) least frequency-tunable and least robust oscillators, and (3) less frequency-tunable yet most robust oscillators. We find that Frequency-tunable networks cannot simultaneously express high robustness, indicating a tradeoff between frequency tunability and robustness.

## I. Introduction

The field of systems biology attempts to understand the dynamics of entire biochemical systems; however, for most systems the complexity of the underlying network of system interactions presents significant barriers to both analytical and computational means of characterizing the dynamics of the system. Therefore, many researchers have noted the usefulness of studying small repeated network topologies termed network motifs as a means to understand systemic dynamics [1,2]. Moreover, it has been shown that certain network motifs are capable of performing specific functions such as switching, adaptation, pattern formation, and oscillation [2,3,4,5,6]. Also, certain networks are not only capable of certain functions but also necessary for them [6].

Oscillation has an important role in biological systems such as circadian rhythms and the eukaryotic cell cycle. In some systems, such as a mammalian heart, it is also advantageous to have oscillation with tunable frequency. John Tyson *et. al.* examined a limited number of network topologies consisting of three nodes and four or fewer edges and identified a few network topologies that are capable of oscillations [2]. Tsai *et. al.* studied three different network motifs, namely a negative feedback loop, coupled negative feedback loops, and coupled positive and negative feedback loops, and found that the interconnected positive and negative feedback loops are capable of changing frequency of oscillation spanning 4-5 orders of magnitude [4].

We consider a set of 249 distinct gene regulatory networks with three nodes and at least one negative feedback loop where the nodes are gene products and the edges are either activation or inhibition of the gene. Each network is transformed into a set of ordinary differential equations similar to the mathematical models in ref. [4]. We searched the parameter space randomly and uniformly to find $10^3$ sets of parameter values per network that produced sustained oscillation. For each set of parameters generating oscillation, we varied a kinetic rate constant between the lower and the upper bifurcation points inside the oscillatory domain and collected about 100 sets of the values of frequency, amplitude, and the kinetic rate constant of that particular oscillation. With those 100 sets, we selected the ranges of frequency and the kinetic rate constant such that the amplitude of oscillation changed less than 50%.

We identified a set of network topologies that can change frequency of oscillation across several orders of magnitude while the amplitude of oscillation remains unchanged over a large range of a kinetic rate constant. Then, we placed 249 networks into a network design space based on their average frequency-tunability and average robustness and classified them into three groups, (a) group I: highly robust, (b) group IV: highly frequency-tunable, (c) group II: least frequency-tunable and least robust oscillators. The common network architectural feature for the networks in the group I is double auto-regulatory positive feedback loops connected by a para-regulatory positive feedback loop; for the networks in the group IV, double auto-regulatory positive feedback loops connected by a para-regulatory negative feedback loop; lastly for the networks in the group II, two auto-regulatory negative feedback loops connected by a para-regulatory positive feedback loop. There is a tradeoff between high frequency-tunability and high robustness, i.e., any networks cannot have both high frequency-tunability and high robustness. We plotted the $10^3$ bifurcation diagrams for each of 249 networks and found that each group has a characteristic ratio of the number of networks exhibiting supercritical Hopf bifurcation to those exhibiting subcritical Hopf bifurcation. Subcritical Hopf bifurcation appears to be related to the high robustness and the high frequency-tunability. Lastly, we performed sensitivity analysis and found that the oscillator is quite robust against the perturbation of the strength of auto-regulatory positive feedback loops whereas it is very fragile to the change of the strength of the backbone negative feedback loop.

**II. Methods and Models**

A. Generation of 249 topologically unique networks

Using the repressilator as the backbone network, we enumerate all possible unique networks having as few as three edges, or as many as nine. The solid edges in figure 1a are inhibitory, and the dashed edges are either activation, inhibition, or no regulation if the edge is absent. Furthermore, edges between nodes are referred to as *para-regulatory* edges, and edges originating and ending on the same node are referred to as *auto-regulatory* edges. *Uniqueness* implies that each network contains a regulation scheme sufficiently different to generate unique dynamics. For example, suppose that upon the backbone network (fig. 1), a para regulatory edge is added such that node C regulates A. If instead the edge had been placed so that node A regulates B, because the backbone itself is symmetric (all edges are inhibitory), whether node C regulates A, or node A regulates B is inconsequential, because both networks generate the same effective dynamics. In this example with one added para-regulatory edge, out the 3 possible networks, the dynamical effect of the edge can be obtained from either one of three permutations, whether the edge extends from node A to B, node B to C, or node C to A; therefore we say there is only 1 *unique* network. In our enumeration method all unique para-regulatory networks are identified, upon which the auto-regulatory edges can be placed and then permuted. Suppose that in addition to the para-regulatory edge added to the backbone network, we add also a single auto-regulatory edge placed either on node A, B, or C. Because of the previous addition of one para-regulatory edge, the symmetry in the backbone network is lost. In this case, all three locations to place the auto-regulatory loop results in a unique network. We applied this counting strategy when zero, one, two, and three para-regulatory edges were added to the backbone network, and then for each number para-regulatory edges, zero, one, two, and three auto-regulatory edges were added. The table below summarizes the result of this counting and categorizes the networks into subgroups according to the number of para and auto-regulatory edges with identical regulation types.

B. Ordinary differential equation model and criterion for oscillation

Each network is expressed as a set of coupled first order, ordinary differential equations following the gene regulatory model of Tsai et al. [4]. Nodes of the network represent gene products and the edges are either activation or inhibition of the gene; all edges are modeled using Michaelis – Menten kinetics. We constructed a set of generalized equations from which any of the 249 unique networks could be generated, by introducing a six component network architecture vector $\eta = \{\eta_{AA}, \eta_{BB}, \eta_{CC}, \eta_{AC}, \eta_{BA}, \eta_{CB}\}$, where each component $\eta_{\alpha\beta}$ can take values {-1, 0, +1}. In this notation, the subscripts are understood as $\alpha$ is regulated by $\beta$. Activation edges are added if $\eta_{\alpha\beta} = +1$, inhibitory edges are added if $\eta_{\alpha\beta} = -1$, and no edge is added when $\eta_{\alpha\beta} = 0$. Auto-regulation occurs is implied if $\alpha = \beta$. By assigning the appropriate values for $\eta_{\alpha\beta}$ for the regulation of each edge in the network, any of the 249 unique networks can be systematically generated. Depending on the network, the ODE system can have as few as 12 parameters or as many as 45 parameters which were sampled uniformly on a linear scale and numerically solved using a fixed stepsize Runge-Kutta 4 (RK4) scheme. Monte Carlo sampling was used to assign values to each random variable in the set of ODEs, which were numerically simulated until $10^3$ parameter sets were collected that produced sustained oscillation. The trajectory of each node was followed for approximately $10^6$ simulation steps (about $10^3$ simulation "seconds") and the sum of the squares of the rates of change of the expression of each node was calculated, (i.e. $\sqrt{\left(\frac{dA}{dt}\right)^2 + \left(\frac{dB}{dt}\right)^2 + \left(\frac{dC}{dt}\right)^2}$). If at any point, the magnitude of this velocity vector dropped below a certain threshold (0.001) the solution would be deemed not oscillatory, otherwise the trajectory was deemed to have *sustained* oscillation. Note that in the generalized ODE system (table 2), the term $k_1(1 - A)$ contains both the constitutive synthesis ($k_1$) and degradation ($-k_1 A$) of species A. Expressing the synthesis and degradation in this form serves to normalize the expression of species $A$ to a maximum of

1. The term $-\frac{k_2 B^{n_1}}{K_1^{n_1}+B^{n_1}} A$ represents the inhibition of species $A$ by species $B$. The signal-response of $A$ to $B$ follows a sigmoidal curve due to the Hill function ($\frac{k_2 B^{n_1}}{K_1^{n_1}+B^{n_1}}$), characteristic of cooperative enzymatic reactions.

C. Binary Search Algorithm to locate an oscillatory domain

For each of the $10^3$ parameter sets yielding sustained oscillation, an oscillatory domain was located by varying the rate constant $k_3$ governing the synthesis and degradation rate of node B until the system could produce sustained oscillation no longer; hence $k_3$ is a bifurcation parameter. Within an interval for $k_3$ bracketed on [0, 1E6], a binary search algorithm was used to locate both upper and lower bounds of *sustained* oscillation. The bracketed interval was determined by testing how high $k_3$ could be varied while maintaining sustained oscillation for a few of the most robust networks. The binary search method tests if a $k_3$ value half-way between the current upper and lower bounds of the interval produces sustained oscillation. Depending upon which bifurcation point is being searched for, the result of the test determines whether the lower or the upper interval bound shall be re-positioned to the location of the test point, and a new half-way point is tested in the next iteration. Each iteration brings the upper and lower bounds of the interval closer together until separation between the upper and lower bound is small enough that the bifurcation point can be approximated; our separation distance was 1E-4. General implementation details of a binary search algorithm are described in ref. [7] and specific details of our implementation can be found in the Supplementary Material.

D. Stepping through the oscillatory domain

The result of the binary search is identification of the upper and lower bounds of the oscillatory domain for a particular parameter set. Out of this oscillatory domain we extract a sub-interval of $k_3$ values that produced oscillation where the amplitude varies less than approximately 50%. This restricted range of kinetic rate constant is found by stepping through the oscillatory domain *adaptively*, substituting each $k_3$ value into the system of ODEs (table 2), re-solving the ODEs, and observing how the amplitudes of oscillation change along the way. Beyond a brief transient period (about 1E4 simulation steps), if the solution fails to reach a steady state during the time needed to collect 4 amplitudes sustained oscillation is determined by $\frac{\Delta Amp_i}{Amp_i} < 0.05, \forall i = \{A, B, C\}$. Amplitudes A, B, and C are defined to be the difference between the maximum and minimum values of each species' respective expression levels.

The size of the step $\Delta k_3$ is chosen to maintain a sufficiently small change in the amplitude of the solution between steps, specifically $\frac{\Delta Amp}{Amp} < 0.05$. If the criteria cannot be met, a minimum stepsize of 1E-4 is used; the maximum stepsize is 1/20$^{th}$ of whole domain. Stepping can be summarized as follows:

$$\text{if } \frac{\Delta \text{Amp}}{\text{Amp}} < tol, \quad \Delta k_{new} = 2\Delta k_{old}$$

$$\text{if } \frac{\Delta \text{Amp}}{\text{Amp}} > tol, \quad \Delta k_{new} = \frac{1}{2}\Delta k_{old}$$

With adaptive stepping, approximately 100 steps are needed to traverse the oscillatory domain, generating 100 sets of the values of frequency, amplitude, and the kinetic rate constant for each of $10^3$ parameter sets collected per network. However, rather than showing the results in terms of the amplitudes of nodes A, B, and C, we present the results using the maximum distance across the three dimensional *limit cycle*. Because the differential equations describing each network are nonlinear (table 2), sustained oscillation of nodes A, B, and C forms a 3-dimensional stable limit cycle when viewed in the state space of the system. We define the distance across the limit cycle to be $\sqrt{\sum_i (\text{Amp}_i)^2}$, for $i = A, B, C$. For any two points on the surface of the limit cycle, the *distance* defines the magnitude of the longest chord. Because distance is a scalar quantity, it is invariant to rotations of the coordinate system or of the limit cycle, and more succinctly communicates the same information as knowing the amplitude of A, B, and C.

E. Determining operational range of frequency and kinetic rate constant for a network

The frequency range over which the limit cycle distance changes less than 50% as the kinetic rate constant is varied is defined as the *operational* frequency range; the corresponding range of the kinetic rate constant is the operational range of kinetic rate constant. Computationally, to identify the operational range of oscillation frequency for a given network, the distance across the limit cycle is plotted against oscillation frequency for all $10^3$ parameter sets; each parameter set is represented as a single curve on this plot (fig. 4, upper panel). We analyzed the data in a logarithmic scale because we are interested only in the order of magnitude of the operational ranges of frequency and kinetic rate constant. Limit cycle distances are scaled by $\log\left(\frac{L.C.\ dist}{\max(L.C.\ dist)}\right)$ and the oscillation frequency is scaled by $\log\left(\frac{Freq}{\max(Freq)}\right)$. Visually, the operational frequency and bifurcation parameter range is associated with the "flat" portion of the curve (fig. 4, upper panel). Quantitatively, the operational ranges are selected only from limit cycle distances greater than -0.3 (fig. 4, lower panel). This selection generates a subset of frequency and parameter ranges whose corresponding limit cycle distances vary less than approximately 50% on an absolute scale. This criterion for selecting operational ranges is applied to all 249 networks, and is valid because the "flat" portion of the curve is associated with the maximum amplitude for most parameter sets in all networks. The order of magnitude of the operational frequency range of each parameter set is given by $\log\left(\frac{\max(Freq)}{\min(Freq)}\right)$ and similarly by $\log\left(\frac{\max(k_3)}{\min(k_3)}\right)$ (fig. 4, lower panel). A parameter set exhibiting a larger operational frequency range is said to be more frequency tunable. Likewise, the greater the operational parameter range is, the more robust is the parameter set against external perturbation. We assign an effective operational range of frequency and kinetic rate constant for the whole network by performing a geometric average over the operational frequency and kinetic rate constant ranges for all $10^3$ parameter sets pertaining to each network motif, providing a measure of how frequency tunable and robust a network is (fig. 2). Figure 2 serves as the metric we use to assess the extent to which a network is frequency tunable and robust.

F. Bifurcation diagrams

Below the lower bound of the oscillatory domain, the network fails to produce sustained oscillation. When $k_3$ assumes the value of the lower bound often the limit cycle distance grows from a very small size up to a maximum distance, and then slowly decreases in size until $k_3$ reaches the upper bound of the oscillatory domain, beyond which the network fails to oscillate. For network having a positive feedback loop, when $k_3$ is evaluated at the lower bound a significant fraction of the parameter sets produce sustained oscillation with limit cycle distances that are at or near their maximum value. For these parameter sets, an initial discontinuous jump is found in the limit cycle distance; figure 5 illustrates bifurcation diagrams for three networks. The vertical axis shows the normalized size of the limit cycle (Log(LC_dist)/Max(LC_dist)), and the horizontal axis shows the normalized range of the bifurcation parameter [Log(Min(k)) – Log(k)]/ [Log(Max(k) ) – Log(Min(k))]. In figure 5, any curve whose limit cycle distance is greater than -0.5 when $k_3$ is evaluated at the lower bound, are assigned the color purple. All curves that have limit cycle sizes less than -0.5 when $k_3$ is evaluated at the lower bound are given the color green. Purple curves are candidates for sub-critical Hopf bifurcation.

G. Naming convention

Any network based upon the cyclic inhibition backbone can be characterized by a six character naming convention. The first three characters refer to those edges that are auto-regulatory, and the latter three characters to those edges that are para-regulatory. Although each edge can take one of three states, activation, inhibition, or no regulation, we distinguish between auto-regulatory and para-regulatory edges. An activation auto-regulatory edge is indicated with the letter "P" for *positive* auto-regulation, inhibitory auto-regulation is indicated by an "N" for *negative* auto-regulation, and no regulation is indicated by "0". Para-regulatory edges that *activate*, are referred to with the letter "A", while those that *inhibit* are referred to by "I", and *no* para-regulation is conveyed by the same use of "0". For both auto-regulation and para-regulation, the first character in each corresponds to the regulation *at* A, the second place corresponds to the regulation *at* B, and the third *at* C. For example, the backbone repressilator is named 000000, a positive auto-regulatory edge on node A and a inhibitory para-regulatory edge from B to C is identified by the name A0000I .

H. Sensitivity Analysis

For all 1024 parameter sets, $k_1$- $k_9$ were varied independently between the lower bifurcation point and the upper bifurcation point. The bracketed region was on [0,1E6]. The term describing the median concentration of substrate at which the velocity of the reaction is one half of the maximum $k_i^{n_i}$ were not varied, nor were the cooperability exponents $n_i$. For the PPP000 network, it appears that the location in the design space is similar so long as only the synthesis and degradation rates are varied ($k_1, k_3, k_5$). The location in design space changes from being frequency tunable and robust to vary robust and not frequency tunable when only the $k_7, k_8, k_9$ rate constants are varied which control the strength of the positive feedback. The location in the design space for networks with $k_7, k_8, k_9$ varied are grouped together. Finally, varying $k_2, k_4, k_6$ led to a decrease in operational frequency and robustness. The rate constants $k_2, k_4, k_6$ regulate the strength of inhibition from the neighbor n the repressilator backbone. In the design space, the networks in which $k_2, k_4, k_6$ were varied also are grouped together.

## III. Results

We consider a set of 249 distinct gene regulatory networks with three nodes and at least one negative feedback loop where the nodes are gene products and the edges are either activation or inhibition of the gene. Each network is transformed into a set of ordinary differential equations similar to the mathematical models in ref. [4]. We searched the parameter space randomly and uniformly to find $10^3$ sets of parameter values per network that produces sustained oscillation. For each set of parameters generating oscillation, we varied a kinetic rate constant between the lower and the upper bifurcation points inside the oscillatory domain and collected about 100 sets of the values of frequency, amplitude, and the kinetic rate constant of that particular oscillation. With those 100 sets, we selected the ranges of frequency and the kinetic rate constant such that the amplitude of oscillation changes less than 50%.

**249 topologically unique networks in network design space.**
We identified a set of network topologies that can change frequency of oscillation across several orders of magnitude while the amplitude of oscillation remains unchanged over a large range of a kinetic rate constant. Then, in Figure 2, we placed 249 networks into a network design space based on their average frequency-tunability and average robustness and classified them into three groups, (a) group I: highly robust, (b) group IV: highly frequency-tunable, (c) group II: least frequency-tunable and least robust oscillators. We make a key observation that the FM tunability of the most robust networks is markedly lower than the FM-tunability of the group-I networks. Conversely, when networks are more FM-tunable as in group-I, their robustness is markedly less than the most robust networks. This observation indicates a tradeoff between frequency and robustness, namely that any network cannot have both high frequency-tunability and high robustness. The existence of the frequency-robustness tradeoff leads us to question what are mechanisms (or architectures) confer the robustness but kills FM and converse?

**Common network topological features.**
Here we identified the common network topological features among the networks in the same group and attributed the dynamical characteristics of the oscillations to the topological features. Each group has about 40-60 networks. The networks in each group were ranked in order of the numerical values of individual dynamical characteristics of the oscillation. One network with highest (or lowest) value per group was designated as a reference network: 0PNA0A network with the highest frequency-tunable range for group I, PPPAA0 network with the highest robustness for group IV, and NN0AAA network with the least frequency-tunable range and the least robustness for group II. Then, we picked other networks one at a time and rotated it until the hemming distance (or the difference) between the reference network and the selected network was minimized. Then, we calculated the average and the standard deviation of network architectural vector elements for each of three groups as presented in Figure 3. This figure depicts both the number of positive and negative feedbacks but also their location on the networks. The average values of network architectural vectors are clearly different between three groups. We drew the "most probable" network topology for each group where the thickness of each edge in the most probable network is determined by the average value of network architectural vector element. The most probable network architecture for the networks in the group I is "PPN0I0" with double auto-regulatory positive feedback loops connected by a para-regulatory positive feedback loop; for the networks in the group IV, "PP00A0" with double auto-regulatory positive feedback loops connected by a para-regulatory negative feedback loop; lastly for the networks in the group II, "NN00I0" with two auto-regulatory negative feedback loops connected by a para-regulatory positive feedback loop. Outstanding difference between two groups I and IV and the group II is the presence of auto-regulatory positive feedbacks. There is a slight topological difference between group I and IV: two auto-regulatory feedbacks are connected by a "certain" negative feedback for group IV while they are connected by a "weakly possible" positive feedbacks. This small topological difference confers high frequency-tunable range to group I and large robustness to group IV.

**Differential behaviors of three most probable networks.**
In Figure 4 (upper panel), we plotted frequency versus the limit cycle distance for each most probable network and average frequency operational range versus average bifurcation parameter operational range (lower panel). In both panel, the group II network is distinctive. The typical shape of the limit cycle distance vs. frequency curve is a narrow inverted "U", implying the frequency range is very narrow. The frequency range is increased for both group I and IV, though the density of curves exhibiting higher frequency variation is greater in group I than in group IV. For a given network, comparing all parameter sets' frequency against the range of kinetic rate constant, the group II network is again most distinctive because the range of kinetic rate giving rise to oscillation is very small. Networks from groups I and IV show more parameter sets with ranges of kinetic rate constant that are much larger; group IV, for example, contains parameter sets that extends up to nine orders of magnitude. The greatest difference in groups I and group IV is the greater density of parameter sets in group IV with large range of kinetic rate constant. Lastly, a common feature of the group I networks is that the distribution of networks is spread more uniformly in range of kinetic rate constant, whereas group IV networks show bimodal distributions with networks having low range of kinetic rate constant and high range of kinetic rate constant.

**Bifurcation mechanisms giving rise to frequency-tunability and the robustness of the oscillation.**

In Figure 5, we plotted the $10^3$ bifurcation diagrams for each of 249 networks and found that each group has a characteristic ratio of the number of supercritical Hopf bifurcation to the number of subcritical Hopf bifurcation. We find that networks in group II have a higher percentage of parameter sets showing a continuous is limit cycle distance up to some maximum and amplitude and subsequent continuous decline of limit cycle distance back to zero (indicative of supercritical Hopf bifurcation). Few group II parameter sets show limit cycle distances which jump discontinuously from 0 to a maximum or near-maximum limit cycle distance, unlike parameter sets from networks in groups I and IV, which have greater than 80% of parameter sets which begin oscillation a near maximal limit cycle distances (indicative of subcritical Hopf bifurcation). On average, most robust networks have about 6 – 8% more parameter sets showing subcritical Hopf bifurcation.

**Oscillators are fragile to the change of the strength of negative feedbacks, but robust to the change of the strength of positive feedbacks.**

We hypothesize that robust oscillators that are not frequency tunable may be resilient to perturbations of positive feedback if the positive feedback contributes most significantly to the dynamics. Similarly, when the effect of the negative regulation loops begins to dominate the dynamics of the network, robustness and frequency tunability will be reduced. Perhaps, the appropriate *balance* of positive feedback and inhibition extends the frequency range of the network while maintaining some robustness. To test this hypothesis, we performed a sensitivity analysis of the PPP000 network to scrutinize the effect of varying kinetic rate constants other than $k_3$ (fig. 6). We found that the oscillator is quite robust against perturbations of the strength of auto-regulatory positive feedback loops, though is very fragile to the change of the strength of the backbone negative feedback loop, in support of the hypothesis. For the PPP000 network, it appears that the location in the design space is similar so long as *only* the synthesis and degradation rates are varied ($k_1, k_3, k_5$). The location in design space changes from being frequency tunable and robust to more robust and less frequency tunable when only the $k_7, k_8, k_9$ rate constants are varied, which control the strength of the positive feedback. We observe that the location in design space for networks with varied $k_7, k_8, k_9$ are grouped together. Lastly, varying $k_2, k_4, k_6$ which regulate the strength of inhibition from the neighbor in the repressilator backbone, led to a decrease in operational frequency and robustness. Similarly, networks with varied $k_2, k_4, k_6$ also are grouped together. More testing is needed to identify what is the proper dynamical balance of positive feedback and negative feedback necessary to increase the frequency of oscillation.

### IV. Conclusion

We identified a set of network topologies that can change frequency of oscillation across several orders of magnitude while the amplitude of oscillation remains unchanged over a large range of a kinetic rate constant. We placed 249 networks into a network design space based on their average frequency-tunability and average robustness and classified them into three groups, (a) group I: highly robust, (b) group IV: highly frequency-tunable, (c) group II: least frequency-tunable and least robust oscillators. The common network architectural feature for the networks in the group I is double auto-regulatory positive feedback loops connected by a para-regulatory positive feedback loop; for the networks in the group IV, double auto-regulatory positive feedback loops connected by a para-regulatory negative feedback loop; lastly for the networks in the group II, two auto-regulatory negative feedback loops connected by a para-regulatory positive feedback loop. There is a tradeoff between high frequency-tunability and high robustness, i.e., any network cannot have both high frequency-tunability and high robustness. We plotted the $10^3$ bifurcation diagrams for each of 249 networks and found that each group has a characteristic ratio of the number of supercritical Hopf bifurcation to the number of subcritical Hopf bifurcation. Subciritcal Hopf bifurcation appears to be related to the high robustness and the high frequency-tunability. Lastly, we performed sensitivity analysis and found that the oscillator is quire robust against the perturbation of the strength of auto-regulatory positive feedback loops whereas it is very fragile to the change of the strength of the backbone negative feedback loop.

IV. Reference

[1] Alon, U, "Network motifs: theory and experimental approaches ", Nature Reviews Genetics, 2007, **8:**450-461.
[2] Tyson, J., et al., "Design Principles of Biochemical Oscillators", Nature Reviews Molecular Cell Biology, 2008, **9:**981-990.
[3] Tyson, J., et al., "Sniffers, buzzers, toggles and blinkers: dynamics of regulatory and signaling pathways in the cell ", Current Opinion in Cell Biology, 2003, **15:**221-231.
[4] Tsai,T., et al., "Robust, Tunable Biological Oscillations from Interlinked Positive and Negative Feedback Loops ", Science, 2008, **321:**126-129.
[5] Ma, W., et al., "Robustness and modular design of the Drosophila segment polarity network ", Molecular Systems Biology, 2006.
[6] Ma, W., et al., "Defining Network Topologies that Can Achieve Biochemical Adaptation" Cell 138, 760–773, August 21, 2009
[7] Knuth, Donald. "The Art of Computer Programming", Volume 3, Sorting and Searching, 3rd edition ed. Addison-Wesley, pp. 409–426.

**Figure captions**

Figure 1: Network consisting of three nodes and of a cyclic negative feedback loop (or repressilator) as a backbone. A solid line denote that the corresponding negative (or inhibitory) edge is present always. A dashed line indicates that the edge can be either positive (+1), negative (-1), or absent (0). Since we don't allow two edges with the same direction, there are only six possible edges at maximum. Thus, the six dimensional (network architectural) vector, η, completely describes the network topology.

Figure 2: 249 topologically unique networks in network design space. Each network is represented by 6 alphabetical letters and placed in an appropriate location, based on its average operational ranges of frequency and robustness. Networks tend to cluster themselves and thus classified into four different groups: group I for large frequency-tunable range (>0.7); group II for small frequency-tunable range (<0.3) and robustness range (<1); group IV for large robustness range (>4.5).

Figure 3: Common network architectures. Top panel: the average and the standard deviation of the edge values of the networks belonging to the same group. The average value close to +1 (-1) indicates the higher probability of having a positive (negative) link; the average value close to zero indicates either the absence of the edge or the equally likely chance of having positive or negative link. Bottom panel: Based on the average edge value, we assign the most probable edge type (either positive or negative link) to each edge whereas the thickness of an edge indicates the average value of that particular edge (or likelihood of the assigned edge type).

Figure 4: Frequency versus limit cycle radius of the oscillations (top panel) and operational frequency-tunable range versus operational range of bifurcation parameter (bottom panel). Each graph contains $10^3$ curves (top panel) or points (bottom panel), which were obtained from the $10^3$ sets of different kinetic rate constants. For top panel, the synthesis rate $K_3$ is varied to obtain the range of the frequency and the limit cycle radius of the oscillation.

Figure 5: Bifurcations. We plotted $10^3$ bifurcation diagrams, each of which has a different set of randomly sampled kinetic rate constants, after normalizing the bifurcation parameter and the radius of the limit cycle. We interpret the continuous increase (or decrease) of the radius of the limit cycle at the onset of oscillation as supercritical Hopf bifurcation whereas the discontinuous jump of the radius of the limit cycle jumps at the onset of the oscillation is regarded as subcritical Hopf bifurcation. There are distinctive differences among three most probable networks as well as other networks in three groups: the high percentage of subcritical Hopf bifurcation is a good indicator of the large frequency-tunable range and the large robustness range.

Figure 6: Sensitivity analysis. Each point in network design space represents the average operational frequency-tunable and robustness ranges of a network (PPP000 in the group I) averaged over $10^3$ sets of randomly sampled kinetic rate constants. Top three points are obtained from perturbation of synthesis rate ($K_1$, $K_3$, $K_5$). A red arrow indicates a shift of the average frequency-tunable and robustness ranges when the strength of auto-regulatory positive feedbacks (kinetic rate constants $K_7$, $K_8$, $K_9$) is perturbed whereas a blue arrow directs the shift induced by that of negative cyclic feedback loop ($K_2$, $K_4$, $K_6$).

Table 1: Enumeration of 249 topologically unique networks

| Number of Identical Para-regulatory edges | Number of Identical Auto-regulatory edges | Number of Unique Motifs |
|---|---|---|
| 1 | 1 | 12 |
| 2 | 1 | 36 |
| 3 | 1 | 6 |
| 1 | 2 | 36 |
| 2 | 2 | 108 |
| 3 | 2 | 18 |
| 1 | 3 | 6 |
| 2 | 3 | 18 |
| 3 | 3 | 9 |
| Grand Total: 249 | | |

Table 2. Generic ODE equation for 249 topologically unique networks: Generalized equations describing any of 249 randomly parameterized network motifs that can be constructed from the repressilator backbone (fig.1). $A$, $B$, and $C$ are the fractions of proteins $A$, $B$, and $C$ that are active; $K$ is the median effective concentration values of the Hill function, and $n$ is the hill coefficient; all $k$'s are kinetic rate constants. Each $\eta_{\alpha\beta}$ can take -1, 0, or +1, for inhibition, no regulation, or activation, respectively. Terms in red describe the repressilator backbone network.

$$\frac{dA}{dt} = k_1(1-A) - \frac{k_2 B^{n_1}}{K_1^{n_1} + B^{n_1}} A + \eta_{AA}\left(\frac{\eta_{AA}+1}{2}\right)k_7(1-A)\left(\frac{A^{n_4}}{K_4^{n_4} + A^{n_4}}\right)$$

$$+ \eta_{AA}\left(\frac{1-\eta_{AA}}{2}\right)k_7 A\left(\frac{A^{n_4}}{K_4^{n_4} + A^{n_4}}\right) + \eta_{AC}\left(\frac{\eta_{AC}+1}{2}\right)k_{10}(1-A)\left(\frac{C^{n_7}}{K_7^{n_7} + C^{n_7}}\right)$$

$$+ \eta_{AC}\left(\frac{1-\eta_{AC}}{2}\right)k_{10} A\left(\frac{C^{n_7}}{K_7^{n_7} + C^{n_7}}\right)$$

$$\frac{dB}{dt} = k_3(1-B) - \frac{k_4 C^{n_2}}{K_2^{n_2} + C^{n_2}} B + \eta_{BB}\left(\frac{\eta_{BB}+1}{2}\right)k_8(1-B)\left(\frac{B^{n_5}}{K_5^{n_5} + B^{n_5}}\right)$$

$$+ \eta_{BB}\left(\frac{1-\eta_{BB}}{2}\right)k_8 B\left(\frac{B^{n_5}}{K_5^{n_5} + B^{n_5}}\right) + \eta_{BA}\left(\frac{\eta_{BA}+1}{2}\right)k_{11}(1-B)\left(\frac{B^{n_8}}{K_8^{n_8} + B^{n_8}}\right)$$

$$+ \eta_{BA}\left(\frac{1-\eta_{BA}}{2}\right)k_{11} B\left(\frac{B^{n_8}}{K_8^{n_8} + B^{n_8}}\right)$$

$$\frac{dC}{dt} = k_5(1-C) - \frac{k_6 A^{n_3}}{K_3^{n_3} + A^{n_3}} C + \eta_{CC}\left(\frac{\eta_{CC}+1}{2}\right)k_9(1-C)\left(\frac{C^{n_6}}{K_6^{n_6} + C^{n_6}}\right)$$

$$+ \eta_{CC}\left(\frac{1-\eta_{CC}}{2}\right)k_9 C\left(\frac{C^{n_6}}{K_6^{n_6} + C^{n_6}}\right) + \eta_{CB}\left(\frac{\eta_{CB}+1}{2}\right)k_{12}(1-C)\left(\frac{C^{n_9}}{K_9^{n_9} + C^9}\right)$$

$$+ \eta_{CB}\left(\frac{1-\eta_{CB}}{2}\right)k_{12} C\left(\frac{C^{n_9}}{K_9^{n_9} + C^9}\right)$$

Table 3. Range of kinetic rate constants for the generic ODE model

| Parameter | Units | Range | Distribution |
|---|---|---|---|
| $k_1, k_3, k_5$ | 1/[time] | 0 - 10 | Uniform |
| $k_2, k_4, k_6, k_7 - k_{12}$ | 1/[time] | 0 - 1000 | Uniform |
| $n_1 - n_9$ | 1 | 1 - 4 | Integer |
| $K_1^{n_1} - K_9^{n_9}$ | Concentration | 0 - 4 | Uniform |

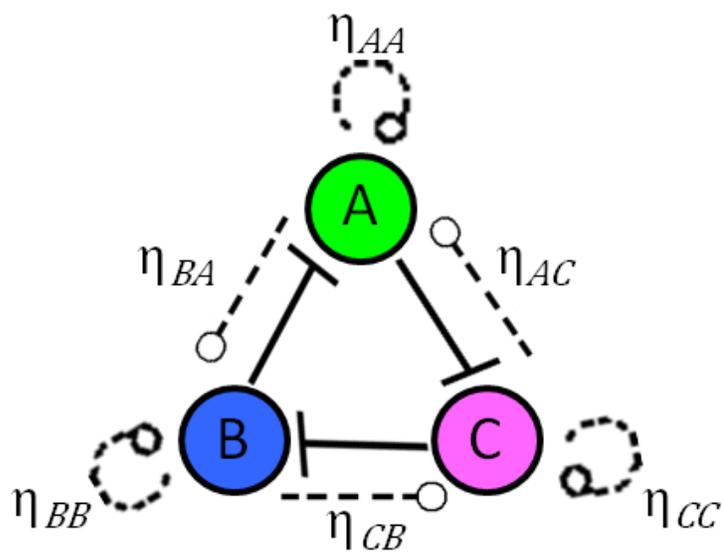

Figure 1

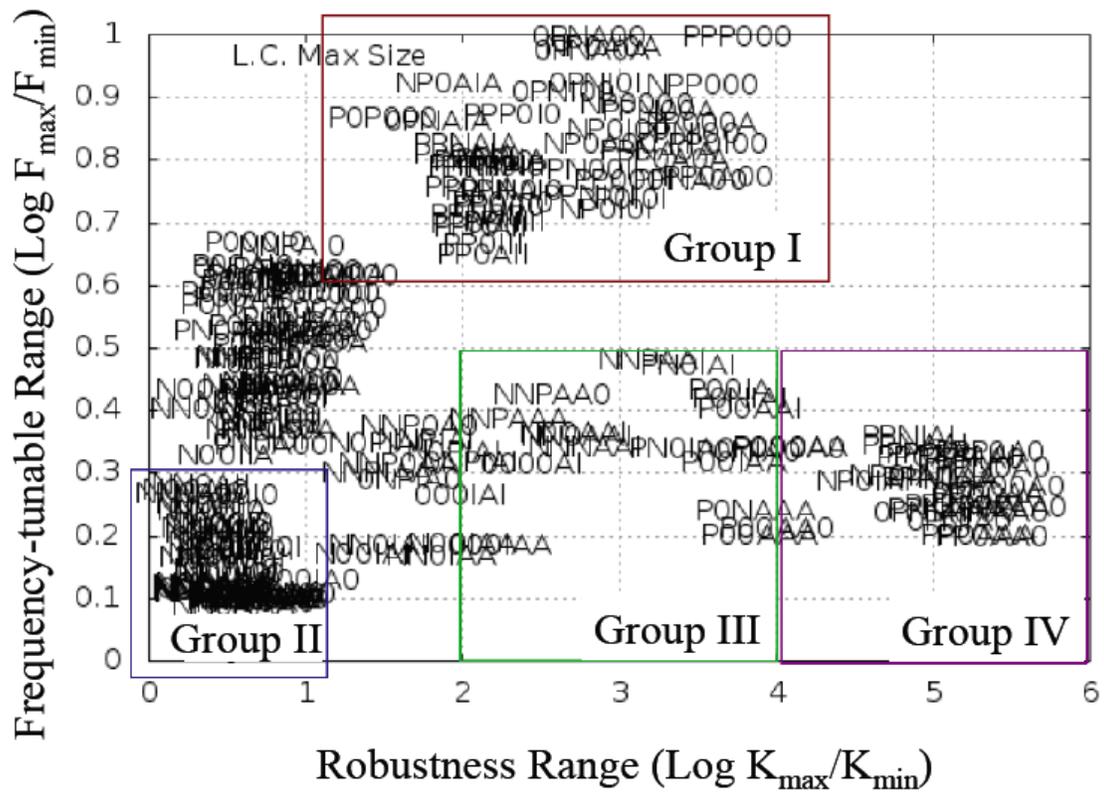

Figure 2

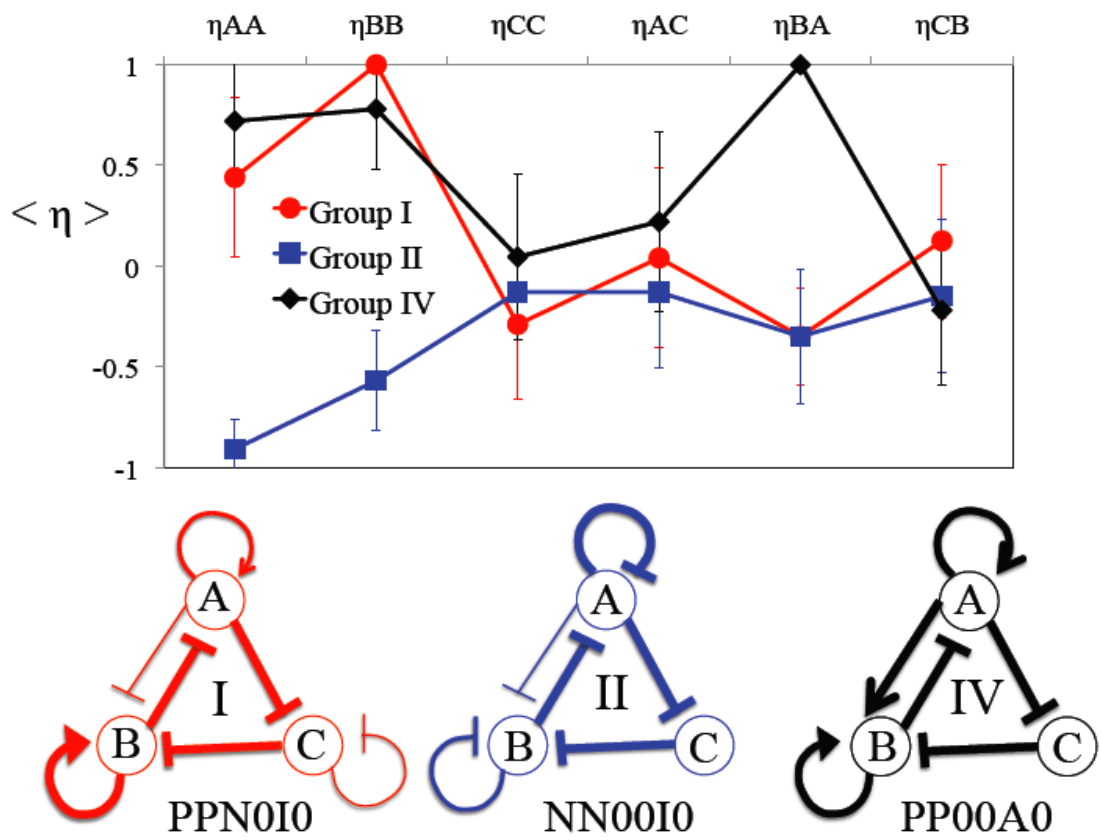

Figure 3

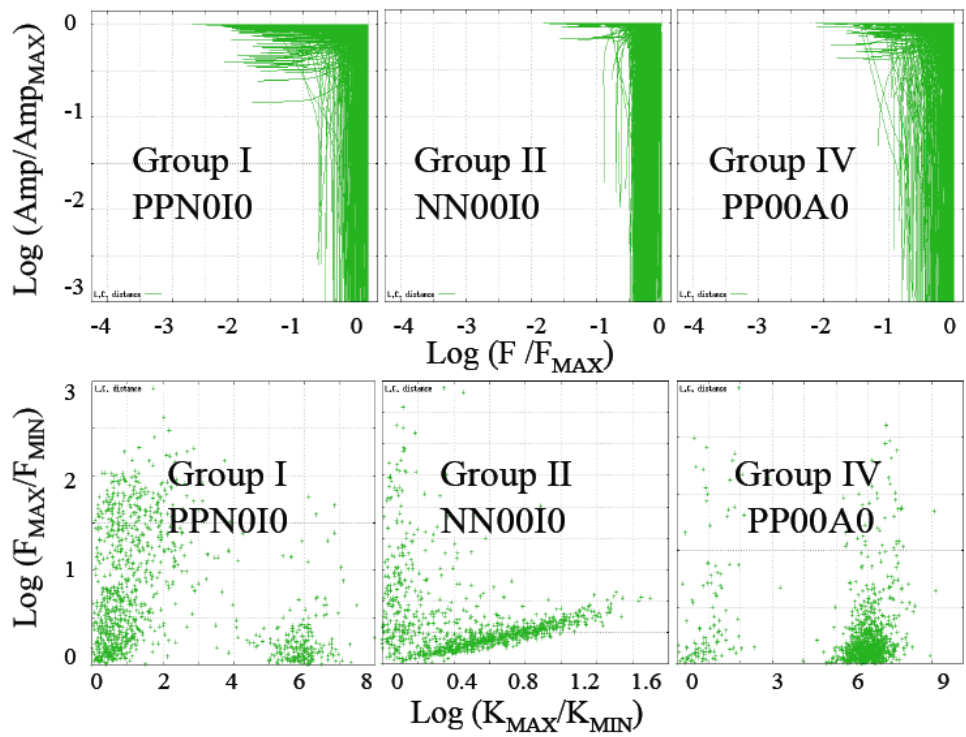

Figure 4

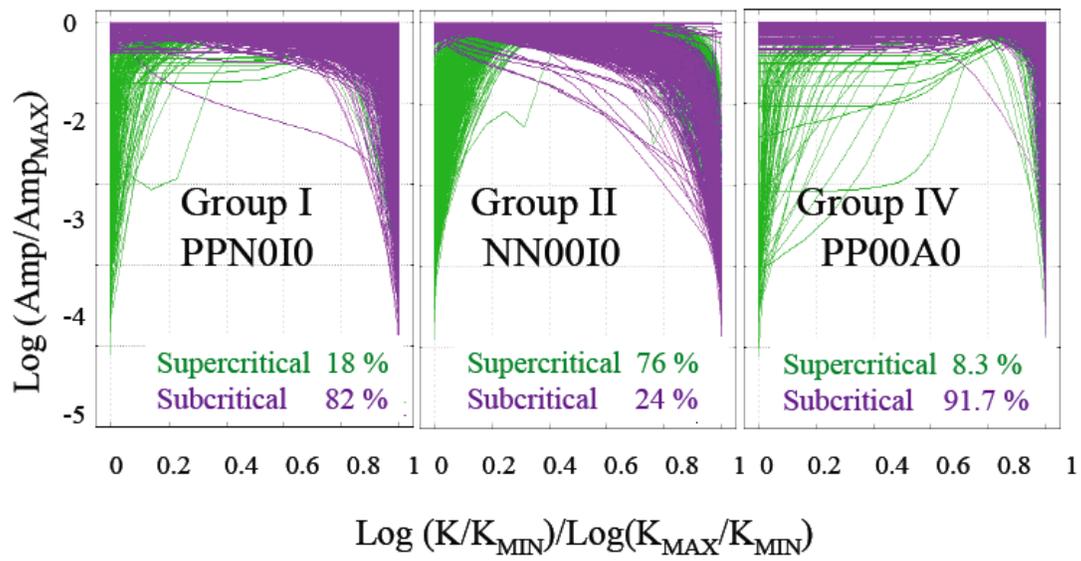

Figure 5

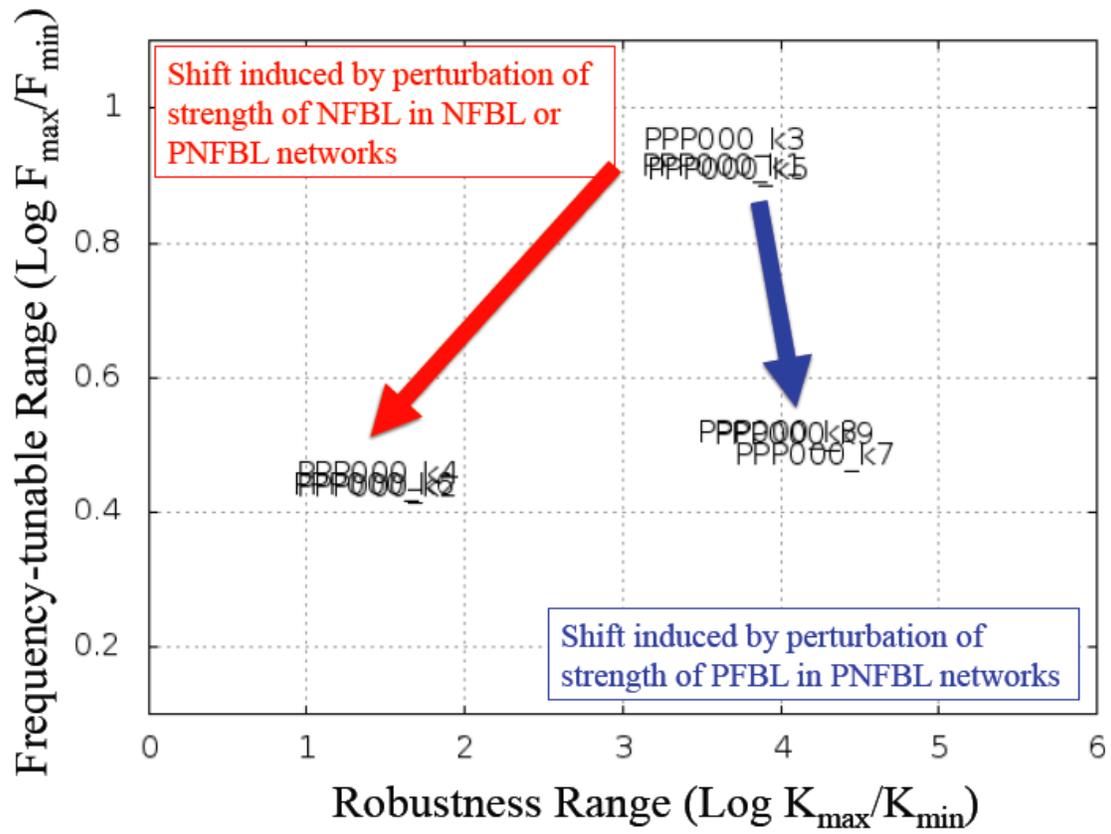

Figure 6